\documentstyle[11pt,fleqn]{article}

\begin{document}

\title{
{\sf QCD  Sum Rules and the $\Pi (1300)$ Resonance}
}
\author{ T.G. STEELE, J.C. BRECKENRIDGE, M. BENMERROUCHE\\
{\sl Department of Physics and Engineering Physics and,}\\
{\sl Saskatchewan Accelerator Laboratory}\\
{\sl University of Saskatchewan}\\
{\sl Saskatoon, Saskatchewan S7N 5C6, Canada.}\\
{V. ELIAS, A.H. FARIBORZ}\\
{\sl Department of Applied Mathematics, 
The University of Western Ontario,} \\
{\sl London, Ontario N6A 5B7, Canada. }
}
\maketitle
\begin{abstract}
Global fits to the shape of the first QCD Laplace sum rule
exhibiting sensitivity to pion-resonance [$\Pi (1300)$] parameters
are performed, leading to predictions for the pion-resonance mass
and decay constant.  
Two scenarios are considered which differ only in their treatment of the
dimension-six quark condensate
$\langle {\cal O}_6\rangle$.  The first scenario assumes an effective 
scale for $\langle {\cal O}_6\rangle$ from other sum-rule applications
which is assumed to be independent of the physical value of the quark mass,
while the second scenario requires self-consistency between the value of
$\langle {\cal O}_6\rangle$ and the current algebra constraint
$2m\langle \bar q q\rangle=-f_\pi^2m_\pi^2$.  Predictions 
of the pion-resonance mass $M_\pi$ and decay constant $F_\pi$ are obtained in these 
two scenarios.
A byproduct of this analysis is 
a prediction of 
the renormalization-group invariant quark mass
$(\hat m_u+\hat m_d)/2$.
\end{abstract}

QCD sum-rule treatments of the pseudoscalar mesons have long been
known \cite{svz} to be successful in predicting  a near-massless pion with a
decay constant $f_{\pi}$ that upholds current algebra (GMOR) constraints
relating the pion parameters $m_{\pi}$ and $f_{\pi}$ to $\langle m
{\bar q} q \rangle $ \cite{gellmann}.
In this approach, the correlation function of charged axial vector currents
$J_\mu^5(x)=\bar u(x)\gamma_\mu\gamma_5 d(x)$ is considered:
\begin{eqnarray}
\Pi_{\mu\nu}(q)& =&i\int d^4x\,e^{iq\cdot x}\,
\langle O\vert T\left(J_\mu^5(x)
J_\nu^{5\dagger}(0)\right)\vert O\rangle
\nonumber\\
&=&\left(g_{\mu\nu}-q_\mu q_\nu/q^2\right)\Pi^T(q^2)
+\frac{q_\mu q_\nu}{q^2}\Pi^L(q^2) \quad .
\label{pilong}
\end{eqnarray}
The longitudinal part $\Pi^L(q^2)$ of this correlation function is related to 
the pseudoscalar resonances with the quantum numbers of the pion.

In the work presented here, we use the 
QCD Laplace sum-rules for 
$\Pi^L(q^2)$
to make explicit predictions
concerning the mass $M_\pi$ and 
the decay constant $F_{\pi}$ of the first excited (pion-resonance) state. 
To leading order in the quark mass, the QCD Laplace sum rules for
$\Pi^L(s)$ and their relation to 
the QCD continuum and
the pseudoscalar resonances in the narrow width approximation 
 are given by \cite{svz,bnry,sr,shuryak,Dorokhov}
\newpage
\begin{eqnarray}
& &
{1\over \pi}\int_0^{\infty} ds \,Im \Pi^L(s)\,e^{-s \tau}=
{\cal R}_0(\tau) 
\\ 
& &
-4 m\langle \bar q q\rangle + O (m^2)  
=2 f_\pi^2 m_\pi^2 e^{-m_\pi^2 \tau} + 2 F_\pi^2 M_\pi^2 e^{-M_\pi^2\tau}
+\ldots
\label{r0phen}
\\
& &
       {1\over \pi}\int_0^{\infty} ds\, Im\Pi^L(s) s \, e^{-s\tau} 
       ={\cal R}_1(\tau)  
\label{r1phen}
\\
& &
 2f_\pi^2 m_\pi^4 e^{-m_\pi^2 \tau} +
        2F_\pi^2 M_\pi^4 e^{-M_\pi^2 \tau}+m^2 c_1(\tau,s0)=
        \nonumber\\
& &\qquad m^2 \Biggl( 
           \frac{3}{2\pi^2\tau^2}\left[
                                        1+\frac{17}{3}\frac{\alpha}{\pi}
                                 \right]
           -\frac{3}{\pi^2\tau^2}\frac{\alpha}{\pi}\left[
                                                         1-\gamma_{_E}
                                                    \right]
-4\,\langle m\bar q q\rangle 
+\frac{1}{2\pi}\langle \alpha G^2\rangle\Biggr.
\nonumber\\
& &\quad
\Biggl.\qquad\qquad\qquad
+\pi\langle{\cal O}_6\rangle\tau
+{3\rho_c^2\over{2 \pi^2\tau^3}} e^{-\rho_c^2 \over {2 \tau}}
\left[ 
       K_0\left( {\rho_c^2\over {2 \tau}} \right) +
       K_1\left( {\rho_c^2\over {2 \tau}} \right)
\right]
\Biggr)
+O\left( m^3 \right)
\label{r1sr}
\\
& &c_1(\tau,s0)=
\frac{3 }{2\pi^2\tau^2}
\Biggl(\left[1+\frac{17}{3}\frac{\alpha}{\pi}\right]
\left[1+s_0\tau\right]e^{-s_0\tau}\Biggl.
\nonumber\\
& &\qquad\qquad\qquad\qquad\qquad
\Biggl.
-2\frac{\alpha}{\pi}\biggl[
e^{-s_0\tau}+E_1(s_0\tau)
+(1+s_0\tau)e^{-s_0\tau}\log(s_0\tau)\biggr]
\Biggr)
\label{continuum}
\end{eqnarray}
where $m=(m_u+m_d)/2$, 
$\langle\bar u u\rangle=\langle \bar d d\rangle\equiv
\langle \bar q q\rangle$,
and the functions $E_1(x)$ and $K_n(x)$ are 
respectively the exponential integral
and modified Bessel functions \cite{as}.
The quantity $m^2c_1(\tau,s_0)$ represents the phenomenological 
contribution of the 
QCD continuum above the continuum threshold 
$s_0$ 
for which perturbative-QCD contributions 
to the longitudinal 
component of the axial vector current correlator
are dual to the phenomenological hadronic contributions 
($s_0 > M_\pi^2$) \cite{bert}. 
We have not included in the hadronic contributions to
${\cal R}_0$ and ${\cal R}_1$ the explicit contribution of the 
three-pion continuum 
of states when $s$ is between $9m_\pi^2$ and $s_0$,  because this 
contribution is
expected to be small compared to the single pion pole.  
For sufficiently small
$s$, the contributions of the pion pole and $3\pi$ continuum to 
the correlation
function $\Pi^L$ can be extracted from their relative contributions to the
pseudoscalar correlation function \cite{pz}
\begin{equation}
\frac{1}{\pi}s\,Im\,\left[\Pi^L(s)\right]_{\pi+3\pi}=
2f_\pi^2 m_\pi^4\left[  \delta(s-m_\pi^2) +
\frac{8s}{3 \left(8\pi f_\pi\right)^4}
\theta\left(s-9m_\pi^2\right)\right]
\label{3pi}
\end{equation}
The relative size of the 
$3\pi$ continuum contribution to ${\cal R}_0$ is easily seen to be negligible, 
as is necessary to ensure the sum-rule validity of the GMOR relation
$f_\pi^2m_\pi^2=-2m\langle \bar q q\rangle$ \cite{gellmann}.  
To ascertain roughly the relative size of these two hadronic contributions to 
${\cal R}_1$, we
can substitute (\ref{3pi}) into the integrand (\ref{r1phen}) and integrate up to 
the continuum threshold $s_0$ [for $s>s_0$, the $3\pi$ continuum and 
all other hadronic effects
are accounted for in $m^2c_1(\tau, s_0)$, the QCD continuum term].   
We find that the $3\pi$ continuum contribution to 
${\cal R}_1$ is less than 7\% of the pion-pole's contribution for
a representative  Borel-parameter choice of $\tau=1 \,{\rm GeV}^{-2}$, 
assuming 
$s_0=4\,{\rm GeV}^2$, and that this percentage is substantially decreased 
if $s_0$ is chosen 
to be smaller ({\it e.g.} 4\% if $s_0=2\,{\rm GeV}^2)$.  
Even if we allow $\tau$ to be as small as
$0.4\,{\rm GeV}^{-2}$ (the minimum value utilized in our fit), and 
$s_0$ to be as large as 
$4\,{\rm GeV}^2$, we still find that the $3\pi$ continuum contribution to 
${\cal R}_1$ is less than
24\% of the pion-pole's contribution, which itself will be seen to be of 
secondary importance 
compared to the pole-contribution  of the first pion excitation state for 
such smaller values of $\tau$.

Finally, the quantity $\langle{\cal O}_6\rangle$ in (\ref{r1sr}) 
denotes the dimension-six quark condensates
\begin{eqnarray}
\langle{\cal O}_6\rangle&\equiv& \alpha_s
\biggl[ 
\left(2\langle \bar u \sigma_{\mu\nu}\gamma_5
T^au\bar u \sigma^{\mu\nu}\gamma_5T^a u\rangle
+ u\rightarrow d\right)
 -4\langle \bar u \sigma_{\mu\nu}\gamma_5T^au\bar d 
 \sigma^{\mu\nu}\gamma_5 T^a d\rangle
\biggr. \nonumber\\
& &\qquad\qquad 
\biggl.+\frac{2}{3}
\langle \left(   
\bar u \gamma_\mu T^a u+\bar d \gamma_\mu T^a d 
\right)
\sum_{u,d,s}\bar q \gamma^\mu T^aq
\rangle\biggr]
\label{o6}
\end{eqnarray}
The  $SU(2)$ breaking effects in (\ref{r1sr}, \ref{continuum})
proportional to $m_u-m_d$
are subleading in the quark mass since
they are proportional to $(m_u-m_d)^2(m_u+m_d)^2$ \cite{bnry}.
As first noted in \cite{svz},  effects other than 
perturbative and power-law terms 
can be present in the ${\cal R}_1$ sum-rule.
Direct instanton contributions to the ${\cal R}_1$ sum-rule
in the instanton liquid model 
\cite{shuryak}  are scaled
by the parameter $\rho_c$, and
can be excised
simply by going to the $\rho_c \rightarrow \infty$ limit.
It should be noted that a sum-rule analysis containing {\em both} 
the rather large two-loop
perturbative term  and direct instanton effects has not previously been 
performed.

The above sum-rules satisfy a renormalization group (RG) equation 
in $\tau$ which implies that
$m$ and $\alpha$ are the running mass and coupling constant at the 
energy scale 
$\tau$ \cite{nar} 
which to next-to-leading order are  \cite{tar}
\begin{eqnarray}
& &\alpha(\tau)=\alpha^{(2)}(\tau)
\left[ 1-\frac{\beta_2}{\beta_1\pi}\alpha^{(2)}(\tau)
\log{\left(-\log(\tau\Lambda^2)\right)}\right]
\label{runalpha}\\
& &\alpha^{(2)}(\tau)=\frac{2\pi}{\beta_1\log(\tau\Lambda^2)}\\
& & m(\tau)\equiv \hat m w(\tau)
\label{runmw}
\\
& &w(\tau)=
\frac{1}{\left[-\frac{1}{2}\log(\tau\Lambda^2)\right]^{-\gamma_1/\beta_1} }
\left[
1-\frac{\gamma_2-\frac{\gamma_1\beta_2}{\beta_1}}{\beta_1^2
\frac{1}{2}\log(\tau\Lambda^2)}
+\frac{\gamma_1\beta_2\log\left[-\log(\tau\Lambda^2)\right]}{\beta_1^3
\frac{1}{2}\log(\tau\Lambda^2)}\right]
\label{runm}\\
& &\beta_1=-\frac{9}{2}\quad ,\quad \beta_2=-8
\quad ,\quad\gamma_1=2\quad ,\quad \gamma_2=\frac{91}{12}
\end{eqnarray}
The quantity $\hat m$ is RG invariant and is thus a fundamental quark mass 
parameter in QCD.  The leading-order versions of (\ref{runm}) and
(\ref{runalpha}) are used for the (leading order) power-law and 
instanton corrections in
(\ref{r1sr}).

The qualitative behaviour of the sum-rules ${\cal R}_0(\tau)$ and 
${\cal R}_1(\tau)$
provides significant information about the first pseudoscalar resonance.  
First, if
$\hat m$ is reasonably small, then ${\cal R}_0(\tau)$ is essentially 
independent
of $\tau$, since it is dominated by $\langle m\bar q q\rangle$.  
This then implies that the phenomenological
side of the sum-rule is dominated by a light pseudogoldstone boson, since
$\exp{\left(-m_\pi^2\tau\right)}\approx 1$ for 
appropriate mass scales [recall $\tau = 1/M^2$ and note that $\Lambda
< M < {\sqrt s_0} $].  
Thus ${\cal R}_0(\tau)$ mainly contains the information that the quark 
condensate 
must balance the phenomenological contribution of the pseudogoldstone
pion, resulting in the GMOR \cite{gellmann} relation
$4m\langle \bar q q\rangle =-2f_\pi^2m_\pi^2$, with
$f_\pi=93\,{\rm MeV}$, as noted above.  

Although the pion dominates ${\cal R}_0(\tau)$, the excited state $M_\pi$ 
in ${\cal R}_1(\tau)$ is enhanced relative to the pion by an
additional factor of $M_\pi^2/m_\pi^2$ relative to its contribution
to ${\cal R}_0(\tau) $. 
For the $\Pi(1300)$ resonance
($M_\pi^2/m_\pi^2\approx 100$), we see that even a 1\% contribution from 
the excited state in 
(\ref{r0phen}) corresponds to the excited state's domination of
(\ref{r1sr}).  
In such a case, the excited state would be too strong to be  
absorbed into the QCD continuum, a point which will be discussed in more 
detail below.

To obtain some quantitative understanding of this excited pion resonance
state, it is necessary to reduce the dependence on the relatively uncertain
value of the quark mass \cite{pdg}.  We 
utilize the
explicit RG dependence in (\ref{runmw}) to obtain 
the following expression from
(\ref{r1sr}): 
\begin{eqnarray}
& &\frac{{\cal R}_1(\tau)}{\hat m^2}=w(\tau)^2 \Biggl( 
           \frac{3}{2\pi^2\tau^2}\left[
                                        1+\frac{17}{3}\frac{\alpha}{\pi}
                                 \right]
           -\frac{3}{\pi^2\tau^2}\frac{\alpha}{\pi}\left[
                                                         1-\gamma_{_E}
                                                    \right]
-4\,\langle m\bar q q\rangle 
+\frac{1}{2\pi}\langle \alpha G^2\rangle
                                                    \Biggr.
\nonumber\\
& &\qquad
\qquad\qquad\qquad\qquad
\Biggl.+\pi\langle {\cal O}_6\rangle\tau
+{3\rho_c^2\over{2 \pi^2\tau^3}} e^{-\rho_c^2 \over {2 \tau}}
\left[ 
       K_0\left( {\rho_c^2\over {2 \tau}} \right) +
       K_1\left( {\rho_c^2\over {2 \tau}} \right)
\right]
\Biggr)
\end{eqnarray} 
Any implicit mass dependence in the above expression occurs through the
value of the dimension-six quark condensate.  There are two points 
of view that can 
be taken for the value of $\langle{\cal O}_6\rangle$.
\begin{enumerate}
\item The scale of $\langle{\cal O}_6\rangle$ is an effective 
(chiral-limiting) value set in numerous sum-rule applications which is not
contingent upon a particular value of the quark mass.  
In this case vacuum saturation and an effective $\langle \bar q q\rangle$ 
scale \cite{svz}are used to find
\begin{equation}
\langle{\cal O}_6\rangle=f_{vs}\frac{448}{27}\alpha
\langle \bar q q\bar q q\rangle
=f_{vs}3\times 10^{-3} {\rm GeV}^6 \equiv \langle {\cal O}_6^{(1)}\rangle
\label{o61}
\end{equation}
where $f_{vs}=1$ for exact vacuum saturation.  Larger values 
of effective dimension-six
operators found in \cite{dimsix} imply that $f_{vs}$  could be as 
large as $f_{vs}=2$.

\item In a self-consistent approach, vacuum saturation is imposed at a 
characteristic $1\,{\rm GeV}$ scale to
give
\begin{equation}
\langle{\cal O}_6\rangle=\frac{448}{27}\alpha(1)
\left[\langle \bar q q\rangle(1) \right]^2
=\frac{112\alpha(1) f_\pi^4 m_\pi^4
\left[-\frac{1}{2}\log(\Lambda^2)\right]^{8/9}}{27 \hat m^2}
\equiv \langle {\cal O}_6^{(2)}\rangle
\label{o62}
\end{equation}
where the GMOR relation $2m\langle \bar q q\rangle=-f_\pi^2 m_\pi^2$  
has been imposed
to make the value of the quark condensate consistent with the quark mass.

\end{enumerate}

By comparing the values of $\langle {\cal O}_6\rangle$ in the two 
scenarios we see that
they are identical if the following identification is made:
\begin{equation}
f_{vs}=\frac{2.6\times 10^{-5} \,{\rm GeV}^2}{\hat m^2}
\end{equation}
Hence a wide enough variation in the parameter $f_{vs}$ in the 
first scenario can account for any
inconsistency between the value of $\langle {\cal O}_6\rangle$, 
the quark mass, and
the GMOR relation.  Furthermore, 
Figure 1 shows that 
the actual numerical effect of these two values of
$\langle {\cal O}_6\rangle$ on the sum-rule ${\cal R}_1/\hat m^2$ 
is small enough to be 
accommodated within the theoretical uncertainties associated with  
${\cal R}_1$
discussed below.  For now it suffices to observe that a wide range 
of smooth functions lying between the
extreme solid curves in the figure occur within our error model, 
accommodating
the small difference between the curves obtained for the two scenarios, 
including variations in 
$\hat m$ for scenario 2.  Although we could accommodate scenario 2 
within the theoretical uncertainties of scenario 1,
we will present explicit results for the two  scenarios distinguished 
by the values
$\langle {\cal O}_6^{(1)}\rangle$ and $\langle {\cal O}_6^{(2)}\rangle$, 
and it 
will be seen that the two are self-consistent when theoretical uncertainties 
in the predicted parameters are considered.

In the first scenario, 
we thus find the following relation between  the QCD  sum-rule and 
phenomenology
by using (\ref{o61}) explicitly, and by 
dividing both sides of (\ref{r1sr}) by $\hat m^2$.
\begin{eqnarray}
& &w(\tau)^2 \Biggl( 
           \frac{3}{2\pi^2\tau^2}\left[
                                        1+\frac{17}{3}\frac{\alpha}{\pi}
                                 \right]
           -\frac{3}{\pi^2\tau^2}\frac{\alpha}{\pi}\left[
                                                         1-\gamma_{_E}
                                                    \right]
-4\,\langle m\bar q q\rangle 
+\frac{1}{2\pi}\langle \alpha G^2\rangle
                                                    \Biggr.
\nonumber\\
& &\qquad\qquad\qquad\qquad
\Biggl.
+f_{vs}3\times 10^{-3}\,{\rm GeV}^6\tau
+{3\rho_c^2\over{2 \pi^2\tau^3}} e^{-\rho_c^2 \over {2 \tau}}
\left[ 
       K_0\left( {\rho_c^2\over {2 \tau}} \right) +
       K_1\left( {\rho_c^2\over {2 \tau}} \right)
\right]
\Biggr)
\nonumber\\
& &\qquad\qquad\qquad\qquad \qquad\qquad\qquad
=\frac{2f_\pi^2m_\pi^4}{\hat m^2}
\left[1+\frac{F_\pi^2M_\pi^4}{f_\pi^2 m_\pi^4}
e^{-M_\pi^2\tau}\right]+w^2(\tau) c_1(\tau,s_0)
\label{r1fit}
\end{eqnarray}
In the first scenario, all $\hat m$ dependence occurs on the right-hand side 
of (\ref{r1fit}).
Input of the  QCD parameters 
permits a least-squares fit 
of the $\tau$-dependence of ${\cal R}_1 (\tau) /{\hat m}^2 $ to
the form 
\begin{equation}
\frac{2f_\pi^2 m_\pi^4}{\hat m^2}\left[ 1+r e^{-M_\pi^2 \tau} \right]
+ w^2(\tau)c_1(\tau, s_0)
\label{lsfit1}
\end{equation}
with the fitted parameters $\hat m$, $r$, $M_{\pi}$ and $s_0$ 
in correspondence
with the right-hand side of (\ref{r1fit}): 
$ r=F_{\pi}^2
M_{\pi}^4/(f_{\pi}^2 m_{\pi}^4)\,\,
$.

In the second scenario, we find a similar relation between the QCD sum-rule
and phenomenology after making use of (\ref{o62}). 
\begin{eqnarray}
& &w(\tau)^2 \Biggl( 
           \frac{3}{2\pi^2\tau^2}\left[
                                        1+\frac{17}{3}\frac{\alpha}{\pi}
                                 \right]
           -\frac{3}{\pi^2\tau^2}\frac{\alpha}{\pi}\left[
                                                         1-\gamma_{_E}
                                                    \right]
-4\,\langle m\bar q q\rangle 
+\frac{1}{2\pi}\langle \alpha G^2\rangle
                                                    \Biggr.
\nonumber\\
& &\qquad
\Biggl.
+{3\rho_c^2\over{2 \pi^2\tau^3}} e^{-\rho_c^2 \over {2 \tau}}
\left[ 
       K_0\left( {\rho_c^2\over {2 \tau}} \right) +
       K_1\left( {\rho_c^2\over {2 \tau}} \right)
\right]
\Biggr)
\label{r1fit2}\\
& &\qquad\qquad
=- \frac{112\alpha(1) f_\pi^4 m_\pi^4
\left[-\frac{1}{2}\log(\Lambda^2)\right]^{8/9}}{27 }
w(\tau)^2\tau\frac{1}{\hat m^2}
+\frac{2f_\pi^2m_\pi^4}{\hat m^2}
\left[1+\frac{F_\pi^2M_\pi^4}{f_\pi^2 m_\pi^4}
e^{-M_\pi^2\tau}\right]+w^2(\tau) c_1(\tau,s_0)
\nonumber
\end{eqnarray}
All the $\hat m$ dependence is again arranged to be  on the 
right-hand side, and the 
parameters $\hat m$, $r$, $M_\pi$ and $s_0$ can again be 
obtained from a least-squares fit.

To obtain these resonance and quark mass 
parameters from a fit to the above shape dependence, we use the 
standard set of values 
$\langle \alpha G^2 \rangle = 0.045 \,{\rm GeV}^4$, 
$\Lambda=0.15 \,{\rm GeV}$, and
$\rho_c=1/600 \,{\rm MeV}$, 
along with the GMOR relation $\langle m \bar q q \rangle = -
f_{\pi}^2 m_{\pi}^2 / 2 $ and physical values for $m_{\pi}$ and
$f_{\pi}$.
The optimum value of the parameters
are then obtained via a fit to the $\tau$ dependence
of the left hand side of (\ref{r1fit}) (scenario 1), or the
left-hand side 
of (\ref{r1fit2}) (scenario 2)
which leads to the 
 smallest value of  a weighted $\chi^2$.  
The weights for the minimum $\chi^2$  are obtained from a 50\% 
uncertainty for
power-law corrections\footnote{The 50\% uncertainty is actually larger  
than the SVZ criterion \cite{svz}
of the (dimensionless) square of the power law corrections.}
and
a 30\% uncertainty for the continuum contributions.  
The magnitude of the relative uncertainty in the power-law, 
perturbative and continuum
contributions is shown in Figure 2 for a typical
value of $s_0$.  The $\tau$ region chosen for the $\chi^2$ 
minimization is the
range for which the relative uncertainty reaches the 20\% level, 
in this case
 $0.4\, {\rm GeV}^{-2} <\tau < 2.5\,{\rm GeV}^{-2}$ .

It is interesting to observe that the  minimum $\chi^2$ 
{\em increases} by an order of magnitude
when the pion-resonance is excluded from the phenomenological model on the
right hand side of (\ref{r1fit}) .
This increase in $\chi^2$ occurs for a variety of scenarios where the 
$3\pi$ continuum is also included with the QCD continuum and pion pole, 
a clear indication that the excited state is too strong to 
be absorbed into  continuum effects.  
The sharp fall-off of ${\cal R}_1$ with increasing
$\tau$ [Fig. 1] also provides strong evidence for a 
substantial pion-resonance contribution, which is 
expected   from (\ref{lsfit1}) to fall off exponentially 
with $\tau$ compared with the constant 
contribution anticipated from the pion itself.  Comparison of the 
form of (\ref{lsfit1}) 
to Fig. 1 indicates that the pion-resonance contribution dominates the 
small $\tau$ region,
whereas the constant contribution from the pion itself is evident in the 
flattening out of ${\cal R}_1$ at large $\tau$.

Uncertainties in the fitted parameters 
 are obtained from
 a Monte-Carlo simulation \cite{lein} 
 based upon a 15\% variation in $\rho_c$, and a simulation of 
 the previously described
power-law and continuum uncertainties.  
In scenario 1, the parameter $f_{vs}$ is allowed to vary in the range
$0<f_{vs}<2$ to accommodate any inconsistency between 
$\langle {\cal O}_6^{(1)}\rangle$
and the fitted value of $\hat m$.
Figure 3 shows the effect
of the 15\% variation in $\rho_c$ on the instanton contributions 
to ${\cal R}_1$.
As evident from the figure, this variation can easily accommodate any
uncertainty associated with the zero-mode approximation 
in Shuryak's instanton liquid model 
\cite{Dorokhov}.
Figure 4 shows that the power-law and continuum uncertainties are 
well described by the empirical formula
\begin{equation}
\delta {\cal R}_1(\tau)/\hat m^2+w^2\delta c_1(\tau, s_0)
=\frac{\sigma_1}{\tau^3}+\sigma_2 \sqrt{\tau}
\label{errmod}
\end{equation}
where $\sigma_1=0.007$ and $\sigma_2=0.006$ in GeV units.
It is now possible to simulate the uncertainties of the minimum $\chi^2$
parameters by performing a Monte Carlo simulation 
with random variations in the parameter set $\rho_c$, $\sigma_1$,  
$\sigma_2$, and $f_{vs}$ (note that $f_{vs}$ only occurs in scenario 1).
\footnote{Random variations in $\sigma_1$ and $\sigma_2$ are confined to
the region $-0.007 <\sigma_1<0.007$ and $-0.006<\sigma_2<0.006$ so that
the simulation does not exceed the error estimate $\delta{\cal R}_1$.}

The results for the parameters and their error estimates in each scenario
are summarized in Table \ref{fittab}.  
The predictions from the two scenarios exhibit an overlap within 
their uncertainties,
so we conclude that the two scenarios for the operator 
$\langle {\cal O}_6\rangle$ are
consistent.
The excellent quality of the minimum $\chi^2$ fits is 
illustrated in Figure 5.  
While the decay constant $F_\pi$ (and hence $r$) has not been measured, 
the PDG
estimate of the $\Pi(1300)$ mass is $1300\pm 100\,{\rm MeV}$ \cite{pdg}.
Our results tend to favour a pion resonance mass at the lower end of the 
experimental
range. 
As is evident from Table 1, the results we obtain are 
quite insensitive to the fitted value 
of $s_0$--- both scenarios remain within 90\% confidence 
levels  for values of
$s_0$ between $1.7\,{\rm GeV}^2$ and $4.4\,{\rm GeV}^2$.  
This entire range, however,
is consistent with the methodological constraint that $s_0$ be greater than 
$M_\pi^2+M_\pi\Gamma$--- {\it i.e.} the requirement that the 
first pion excitation 
resonance be entirely below the QCD continuum threshold for 
the resonance not to be
absorbed in the QCD continuum contribution.  Note that the 
Particle Data Guide
\cite{pdg}  estimates $\Gamma$ to be between $200$ and $600\,{\rm MeV}$, 
suggesting
values of $s_0$ in excess of $2\,{\rm GeV}^2$.
Our quark mass estimates for 
$m(1 {\rm GeV})=[m_u({\rm 1 GeV})+m_d({\rm 1 GeV})]/2$
are also consistent with \cite{pdg} values.

The role of direct instanton contributions to the sum-rule in the 
instanton liquid model 
can be
understood by comparing to a fit in which such contributions are
absent (the $\rho _c \rightarrow \infty$ limit). Corresponding parameter
values for scenario 1 in the $\rho_c \rightarrow \infty $ limit are
$M_\pi = 1.34 \pm 0.16 \,{\rm GeV}$,
$r=8.88\pm 3.2$, $\hat m= 12.15\pm 2.0\,{\rm MeV}$
($m(1 GeV)=9.14\pm 1.5\,{\rm MeV}$), and $s_0=3.44\pm 1.4 \,{\rm GeV}$.  
Thus, the effect of direct
instanton contributions is to lower both the pion-resonance mass and
its decay constant.

It is also possible to gain some insight into the instanton size 
$\rho_c$ by adding $\rho_c$ to the set of fit parameters.  
We then find an optimum  
value of $\rho=1.47 \,{\rm GeV}^{-1}$, $\hat m=10.2 \,{\rm MeV}$,
$r=6.68$, $M_\pi=1.13\,{\rm GeV}$ and $s_0=3.8\,{\rm GeV}^2$.  
A Monte Carlo simulation of errors for this case is beyond our present 
computational capacity.

\noindent
{\bf Acknowledgements:}  TGS and VE are grateful for the
financial support of the Natural Sciences and Engineering Research Council of
Canada (NSERC).

\newpage
\begin{table}
\caption{Pion resonance and quark mass parameters obtained from 
minimization of the weighted $\chi^2$.
The two scenarios correspond to the two possibilities for the
value of $\langle {\cal O}_6\rangle$.  All uncertainties are at 
the 90\% confidence level.}  
\label{fittab}
\vspace{1.0cm}
\begin{tabular}{||l|c|c||}\hline\hline
Scenario  &        1   &     2 \\ \hline \hline
$M_\pi$    &  $0.996\pm 0.25\,({\rm GeV})$ & $0.95\pm 0.24\,({\rm GeV})$ 
\\ \hline
$r$              & $5.40\pm 3.8$   & $7.28\pm 5.2$   \\  \hline
$\hat m$         & $10.88\pm 2.9\,({\rm MeV})$   &  
$13.95 \pm 3.8\,({\rm MeV})$ \\ \hline
$m(1\,{\rm GeV})\,$  & $8.03\pm 2.2\,({\rm MeV})$  
& $10.30\pm 2.8\,({\rm MeV})$\\ \hline
$s_0$ & $3.21\pm 1.5\,({\rm GeV}^2)$ & $3.02\pm 1.4 \,({\rm GeV}^2)$ 
\\ \hline\hline
\end{tabular}

\end{table}

\newpage

\newpage

\noindent{\large{\bf Figure Captions} }

\begin{itemize}

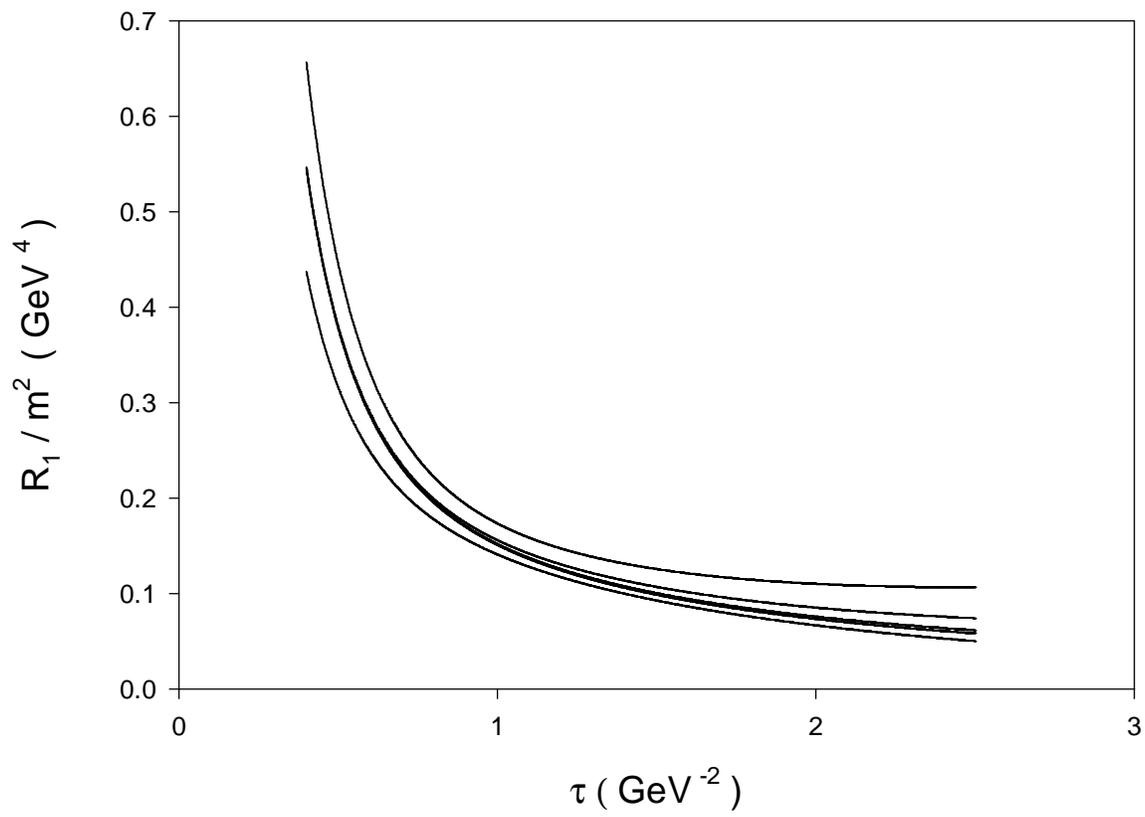
\item[{\bf Figure 1:}] Upper and lower curves  show extremes of the error
model in scenario 1.  The middle solid curves show the dependence of the
sum-rule ${\cal R}_1(\tau)$ on $\langle {\cal O}_6\rangle$ for both 
scenarios, including quark mass dependence in scenario 2 for 
$5\,{\rm MeV}<\hat m<20\,{\rm MeV}$.

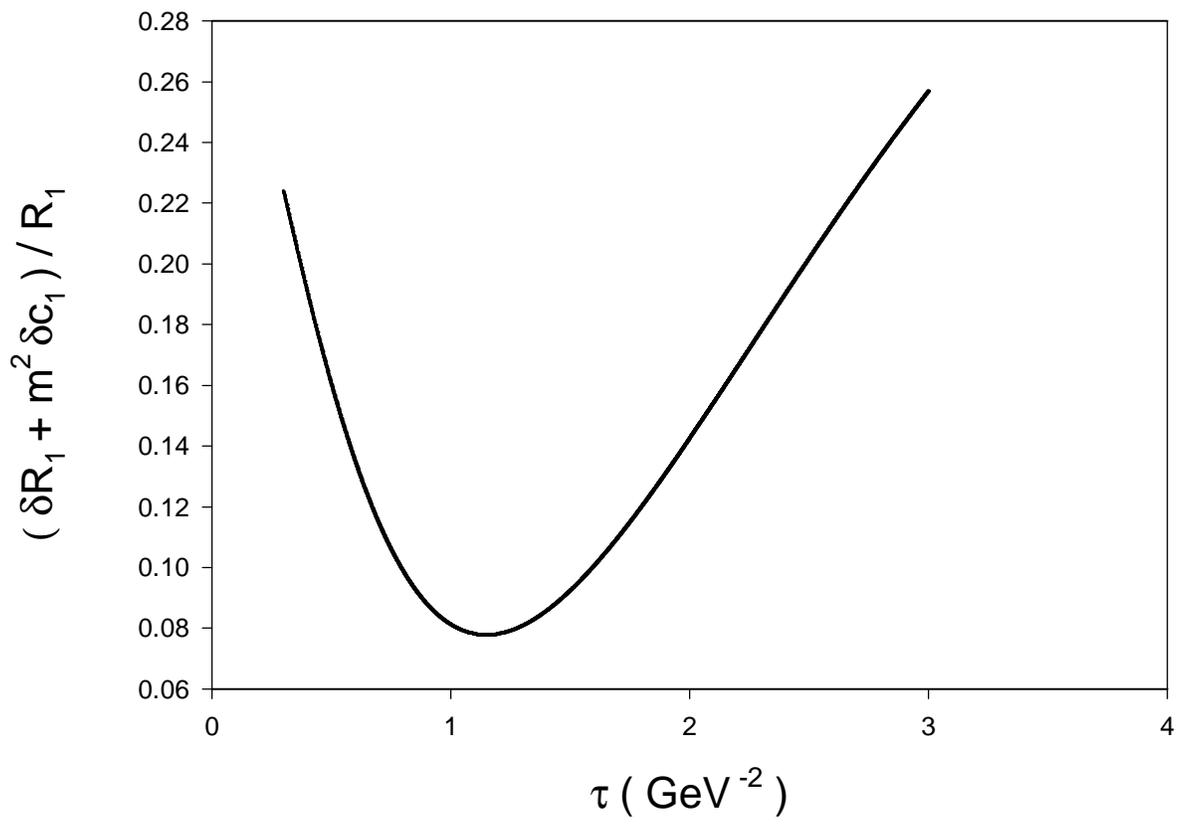
\item[{\bf Figure 2:}]
Relative uncertainty in the power-law, perturbative, and 
continuum contributions to\\ 
${\cal R}_1(\tau)/\hat m^2-w^2(\tau) c_1(\tau,s_0)$ for $s_0=3\,{\rm GeV}^2$.

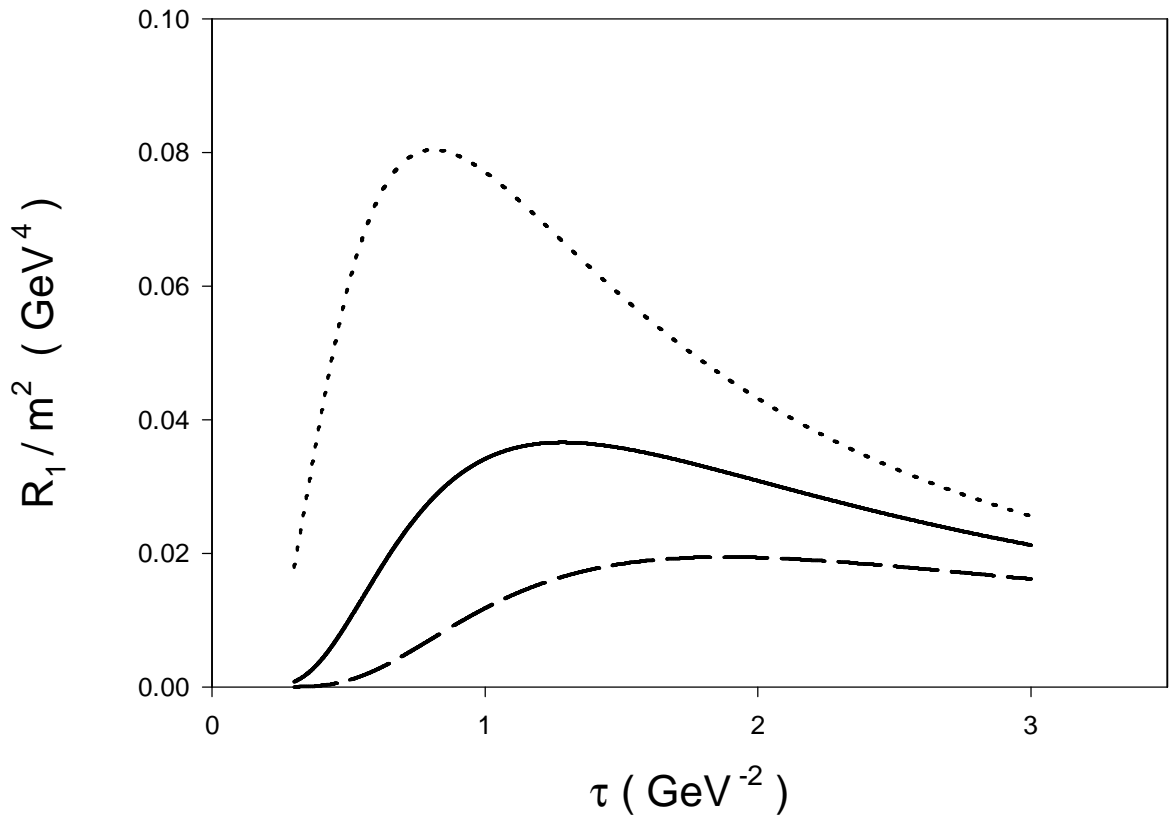
\item[{\bf Figure 3:}]
Instanton contributions to ${\cal R}_1/\hat m^2$ for 
$\rho_c=1/600\,{\rm MeV}$ are shown in the central curve.
The upper (lower) curves are the instanton contributions for 
$\rho$ 20\% smaller (larger) than the central value of 
$\rho_c=1/600\,{\rm MeV}$.

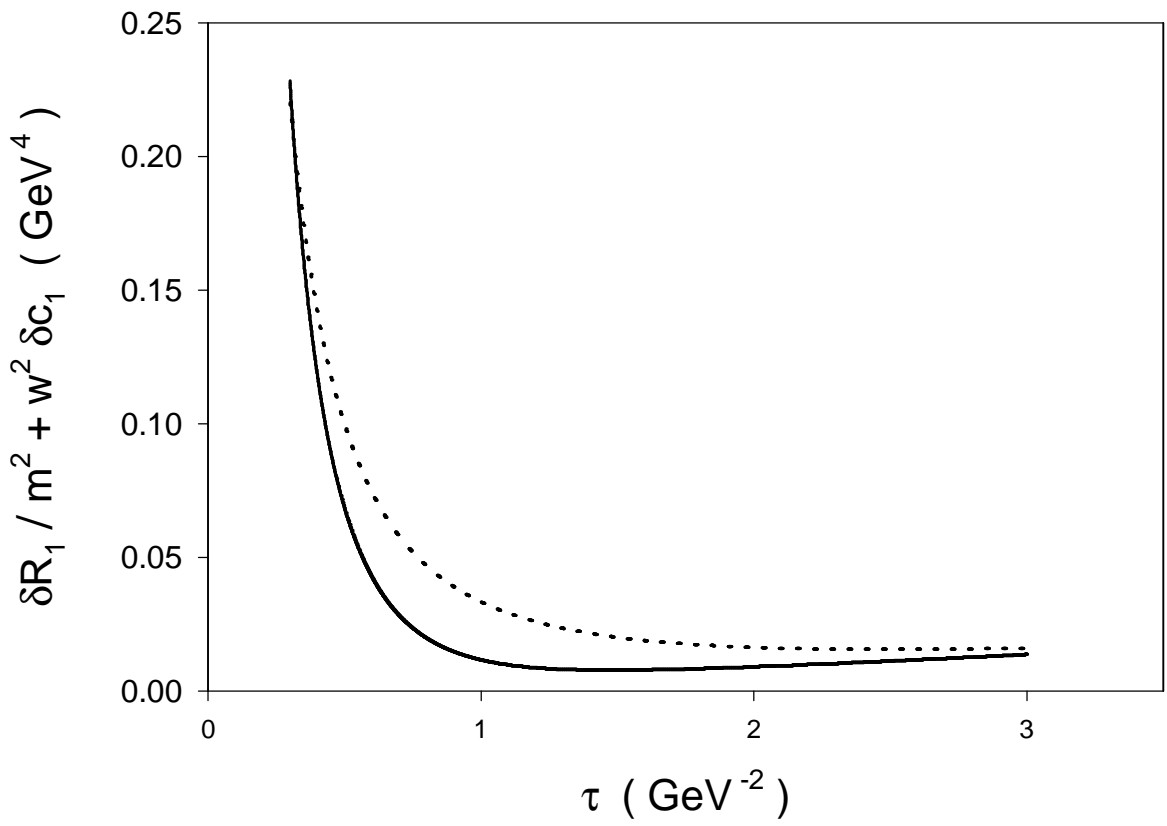
\item[{\bf Figure 4:}] Error model from (\ref{errmod}) (upper curve) and 
uncertainty 
$\delta {\cal R}_1(\tau)/\hat m^2+w^2(\tau)\delta c_1(\tau, s_0)$ 
for $s_0=3\,{\rm GeV}^2$ (lower curve)

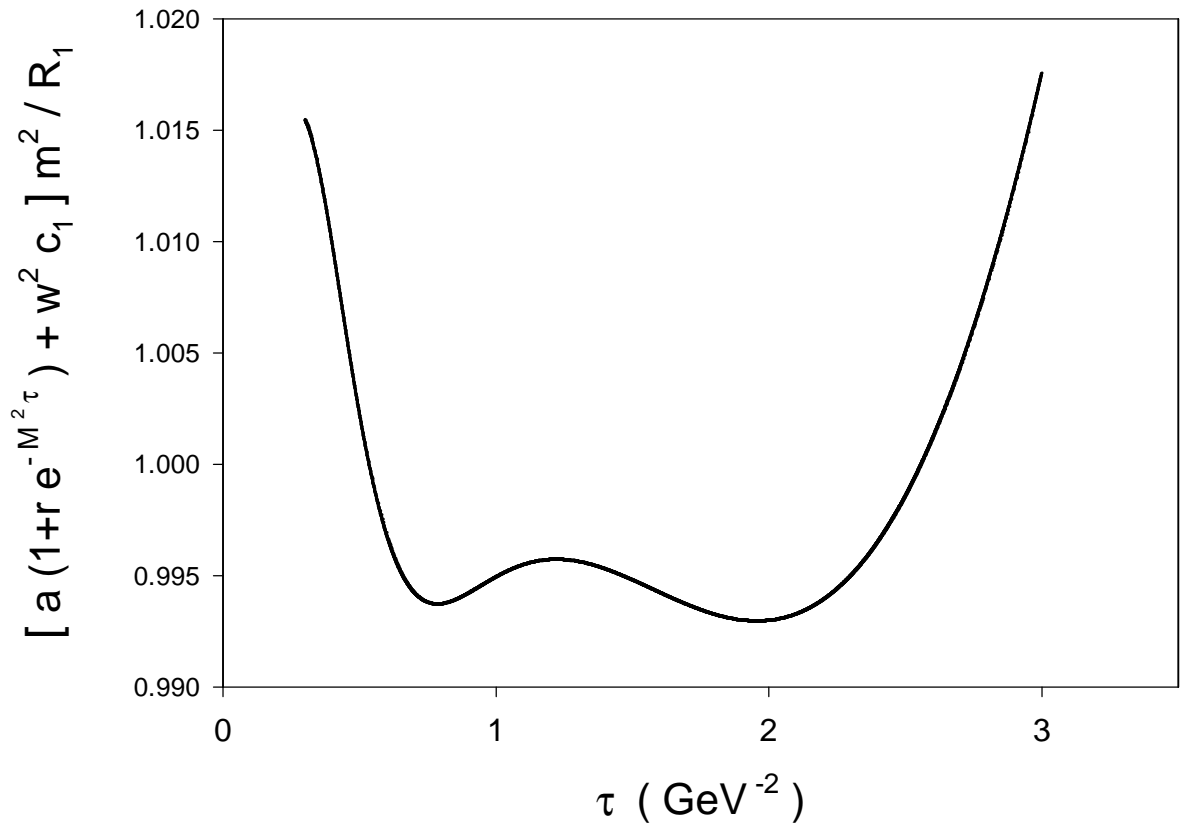
\item[{\bf Figure 5:}] Ratio of the left and right-hand sides 
of (\ref{r1fit}) for
the fitted parameters leading to a minimum weighted $\chi^2$.

\end{itemize}

\end{document}